\documentclass[aps,prb,twocolumn,amsmath,amsfonts,floatfix,showpacs,letterpaper,superscriptaddress]{revtex4}

\usepackage{graphicx}

\DeclareMathOperator{\Tr}{Tr}
\DeclareMathOperator{\curl}{curl}
\DeclareMathOperator{\divv}{div}

\newcommand{\beq}[1]{\begin{equation}\label{#1}}
\newcommand{\eeq}{\end{equation}}
\newcommand{\refeq}[1]{Eq.~(\ref{#1})}
\newcommand{\refeqs}[2]{Eqs.~(\ref{#1})\nobreakdash--(\ref{#2})}
\newcommand{\refeqand}[2]{Eqs.~(\ref{#1}) and (\ref{#2})}

\newcommand{\beqm}[1]{\begin{multline}\label{#1}}

\newcommand{\punc}[1]{\,{\text{#1}}}
\newcommand{\sub}[1]{_{\text{#1}}}

\newcommand{\Kagome}{Kagome}
\newcommand{\kagome}{kagome}

\newcommand{\Z}{{\mathcal{Z}}}
\newcommand{\T}{{\mathcal{T}}}
\newcommand{\Ham}{{\mathcal{H}}}
\newcommand{\Act}{{\mathcal{S}}}
\newcommand{\Lag}{{\mathcal{L}}}

\newcommand{\clH}{\mathcal{E}}
\newcommand{\quH}{\mathcal{H}}

\newcommand{\CubicTrans}{\mathbb{T}}
\newcommand{\KagomeTrans}{\mathbb{K}}
\newcommand{\Rot}{\mathbb{R}}
\newcommand{\RfI}{\mathbb{I}}
\newcommand{\RfX}{\mathbb{X}}

\newcommand{\Qop}{\mathbb{Q}} 
\newcommand{\Qhat}{\hat{Q}} 

\newcommand{\nd}{^{\phantom{\dagger}}}

\newcommand{\Jv}{{\mathbf{J}}}
\newcommand{\Av}{{\mathbf{A}}}
\newcommand{\Avb}{{\bar{\mathbf{A}}}}
\newcommand{\Ab}{{\bar{A}}}

\newcommand{\del}{\boldsymbol{\nabla}}
\newcommand{\phiv}{\boldsymbol{\varphi}}
\newcommand{\sigv}{\boldsymbol{\sigma}}
\newcommand{\mv}{\mathbf{m}}

\newcommand{\Mm}{\mathbf{M}}

\newcommand{\Psiv}{\boldsymbol{\Psi}}
\newcommand{\parv}{\boldsymbol{\partial}}
\newcommand{\kpar}{\tilde{\partial}}
\newcommand{\kparv}{\tilde{\boldsymbol{\partial}}}
\newcommand{\nv}{\mathbf{n}}

\newcommand{\ev}{\mathbf{e}}
\newcommand{\zetav}{\boldsymbol{\zeta}}
\newcommand{\dv}{\mathbf{d}}
\newcommand{\deltav}{\boldsymbol{\delta}}

\newcommand{\kv}{{\mathbf{k}}}

\newcommand{\rv}{{\mathbf{r}}}
\newcommand{\xv}{{\mathbf{x}}}
\newcommand{\bv}{{\mathbf{b}}}

\newcommand{\zerov}{{\boldsymbol{0}}}

\newcommand{\kagx}{\tilde{x}}
\newcommand{\kagy}{\tilde{y}}
\newcommand{\sublattice}[1]{\mathrm{#1}}

\newcommand{\ee}{\mathrm{e}}
\newcommand{\ii}{\mathrm{i}}
\newcommand{\dd}{\mathrm{d}}

\newcommand{\ket}[1]{{|#1\rangle}}
\newcommand{\bra}[1]{{\langle#1|}}

\newcommand{\U}[1]{\ensuremath{\mathrm{U}(#1)}}
\newcommand{\SU}[1]{\ensuremath{\mathrm{SU}(#1)}}
\newcommand{\SO}[1]{\ensuremath{\mathrm{SO}(#1)}}
\renewcommand{\O}[1]{\ensuremath{\mathrm{O}}(#1)}

\newcommand{\NCP}{NC$CP^1$}

\newcommand{\degrees}{{^\circ}}

\newcommand{\putinscaledfigure}[1]{\begin{center}\includegraphics[width=\columnwidth]{#1.eps}\end{center}}

\begin{document}

\title{Classical to quantum mapping for an unconventional phase transition in a three-dimensional classical dimer model}

\author{Stephen Powell}
\affiliation{Rudolf Peierls Centre for Theoretical Physics, University of Oxford, 1 Keble Road, Oxford, OX1 3NP, United Kingdom}
\affiliation{Joint Quantum Institute and Condensed Matter Theory Center, Department of Physics, University of Maryland, College Park, MD 20742, USA}
\author{J. T. Chalker}
\affiliation{Rudolf Peierls Centre for Theoretical Physics, University of Oxford, 1 Keble Road, Oxford, OX1 3NP, United Kingdom}

\begin{abstract}
We study the transition between a Coulomb phase and a dimer crystal observed in numerical simulations of the three-dimensional classical dimer model, by mapping it to a quantum model of bosons in two dimensions. The quantum phase transition that results, from a superfluid to a Mott insulator at fractional filling, belongs to a class that cannot be described within the Landau-Ginzburg-Wilson paradigm. Using a second mapping, to a dual model of vortices, we show that the long-wavelength physics near the transition is described by a \U1 gauge theory with \SU2 matter fields.
\end{abstract}


\maketitle

\section{Introduction}
\label{SecIntroduction}

In the presence of strong local constraints, certain classical systems exhibit so-called `Coulomb phases', where correlation functions have power-law forms and nontrivial directional dependence.\cite{Youngblood,Henley,Huse,Bramwell,Kasteleyn} These phases are of considerable theoretical interest because of their unusual properties, and are also of direct experimental relevance. Examples include frustrated magnets such as spin ice \cite{Bramwell} and molecular dimers adsorbed onto surfaces.\cite{Blunt}

Coulomb phases stand in contrast to ordered phases characterized by a broken symmetry and an associated order parameter.\cite{Landau} Continuous transitions between Coulomb and ordered phases present a novel situation where the behavior at the transition cannot be captured purely in terms of a Ginzburg-Landau theory of the fluctuations of this order parameter. Instead, a complete description of the transition requires the long-range correlations in the Coulomb phase to be taken into account.\cite{Bergman,Alet,Jaubert,SpinIce,Letter,Pickles}

Classical dimer models \cite{Fowler,Kasteleyn} are among the simplest possible model systems that exhibit Coulomb phases, and the discovery of a direct transition into an ordered crystalline phase in the dimer model on the cubic lattice has stimulated considerable interest.\cite{Alet,Misguich,Letter,Charrier,Chen,Papanikolaou} The question of whether the transition is continuous or first order remains controversial, but it is clear that the correlation length at the transition is either divergent or at least several orders of magnitude larger than the lattice spacing. The long-distance properties near the transition can therefore be described in terms of a continuum theory including only the relevant degrees of freedom. It has been suggested\cite{Letter,Charrier,Chen} that the appropriate description is given in terms of a noncompact \U1 gauge theory with \SU2-symmetric matter fields, or noncompact $CP^1$ (\NCP).

In this paper, we analyze the transition in the classical dimer model on the cubic lattice by using a mapping to an equivalent quantum model in two spatial dimensions. A brief outline of this mapping and the predictions that result has been presented previously.\cite{Letter} We use the standard approach of relating classical statistical mechanics in $d$ dimensions to quantum mechanics in $d-1$ dimensions, which in principle provides an exact identity between the partition functions in the two cases.

By an appropriate choice of the mapping, we represent the interacting dimers on the links of a cubic lattice as hard-core bosons on the sites of a \kagome\ lattice. The Coulomb phase then corresponds to the condensed phase of the bosons, and the power-law correlations can be understood in terms of the coupling to the phonon mode of the superfluid. The thermal transition into the dimer crystal is equivalent to a (zero-temperature) quantum phase transition from the superfluid to a Mott insulator at fractional filling.

Interestingly, this belongs to a class of unconventional quantum phase transitions considered by Balents et al.\cite{Balents} In these cases, the phases on the two sides of the transition have different order parameters, and a na\"\i ve application of the Landau-Ginzburg-Wilson (LGW) paradigm predicts that a continuous transition requires simultaneous fine-tuning of two independent parameters. Balents et al.\cite{Balents}\ instead propose a critical theory in terms of dual vortex degrees of freedom, which allows for a generic continuous transition between the two phases. Applying this approach to our effective quantum model gives a continuum gauge theory for the transition that is the same as has been obtained using a direct mapping carried out in three spatial dimensions.\cite{Charrier,Chen}

We have previously\cite{SpinIce} applied a similar approach to a model of nearest-neighbor spin ice, where a transition takes place from a high-temperature Coulomb phase to a low-temperature saturated phase, which has neither power-law correlations nor symmetry breaking.\cite{Jaubert} The mapping again leads to a theory of quantum bosons, but in that case the thermal phase transition maps to the standard quantum phase transition between the superfluid and a vacuum state, described by the conventional LGW approach.

In the remainder of this section, we define the cubic dimer model and review its phase structure. In Section~\ref{SecMapping}, we introduce the mapping from the classical dimer model to a quantum model of bosons on the \kagome\ lattice. We then show, in Section~\ref{SecCoulombPhase}, that the properties of the Coulomb phase of the classical model can be understood in terms of those of the condensed phase of the quantum bosons. In Section~\ref{SecTransition}, we address the phase transition and use a dual picture in terms of vortices to derive a continuum theory to describe the critical properties. We conclude in Section~\ref{SecDiscussion} with discussion. In the Appendix, we briefly consider modifications to the dimer model that lead to the appearance of an intermediate disordered phase and show how this can be understood in terms of the quantum mapping.

\subsection{Model}
\label{SecModel}

We treat a model of classical dimers on the links of a simple cubic lattice. In a given configuration of the classical model, each link is occupied by either one dimer or none, with the close-packing constraint that every site of the lattice has a total of precisely one dimer on the adjoining links. We define the variables $d_\mu(\rv) \in \{0,1\}$, giving the number of dimers on the link joining the sites at $\rv$ and $\rv + \deltav_\mu$, where $\deltav_\mu$ ($\mu \in \{x,y,z\}$) is a basis vector of the cubic lattice. The close-packing constraint can then be expressed as
\beq{EqClosePacking}
\sum _\mu \left[ d_\mu(\rv) + d_\mu(\rv - \deltav_\mu)\right] = 1\punc{,}
\eeq
for all sites $\rv$.

Each configuration is assigned an energy (and hence Gibbs-Boltzmann weight) that favors the parallel alignment of the dimers on neighboring links. In the simplest case, the energy of a configuration is $\clH = -n_\parallel$, where $n_\parallel$ is the total number of plaquettes (of any orientation) with parallel dimers. In a system with periodic boundary conditions and an even number of sites in all three directions, the minimum of $\clH$ occurs when all dimers are placed on parallel links, giving $n_\parallel = N$, the total number of sites. There are six such configurations; one example has $d_\mu(\rv) = 1$ for $\mu = z$ and $r_z$ odd, and zero otherwise. We expect our continuum theory to be equally applicable to other potentials with the same symmetry that also favour columnar crystalline order.

At temperature $T = 0$, the system minimizes $\clH$ by selecting one of these six configurations, breaking both the translational and rotational symmetries of the lattice. We define the `magnetization' order parameter
\beq{EqOrderParameter}
m_\mu(\rv) = \frac{1}{2}(-1)^{r_\mu}\left[ d_\mu(\rv) - d_\mu(\rv - \deltav_\mu) \right]\punc{,}
\eeq
so that the six ground states have $\mv \in \{ \pm \deltav_x , \pm \deltav_y , \pm \deltav_z \}$ for all $\rv$. For small positive $T$, a low-temperature expansion predicts that $\langle\mv\rangle$ will remain nonzero and directed along one of the cubic axes.

In the opposite limit, $T \rightarrow \infty$, there is equal statistical weight for all configurations obeying the close-packing constraint, the number of which grows exponentially with $N$. This limit has been considered by Huse et al.,\cite{Huse} who showed that the system exhibits a Coulomb phase, where the correlation functions are algebraic at long distances. In particular, for the connected part of the dimer correlation function, one finds the standard 3D dipolar form,\cite{Huse}
\beq{EqDipolarCorrelations}
\langle d_\mu(\rv) d_\nu(\zerov) \rangle _\mathrm{c} \sim \eta _\rv \frac{3 r_\mu r_\nu - r^2 \delta _{\mu \nu}}{r^5}\punc{,}
\eeq
where $\eta_\rv = (-1)^{\sum_{\mu}\!r_\mu}$ is $\pm 1$ on the two sublattices. This form for the correlation functions is expected to persist for large finite temperatures.

The high- and low-temperature phases cannot be smoothly connected and must therefore be separated by one or more phase transitions. High-precision Monte Carlo simulations show that there is in fact a single phase transition at a critical temperature\cite{Alet} $T\sub{C} \approx 1.675$, and show that this is either continuous or very weakly first order. In either case, the correlation length at the transition is much larger than the lattice spacing and so a continuum description should be expected to capture the long-wavelength properties near the transition.

As we have noted, we expect our theory to be equally applicable in the presence of modifications that maintain the symmetry of the configuration energy and the ordered states. One can also consider modifications of the model that reduce the cubic symmetry of the lattice, thereby reducing the symmetry of the effective quantum Hamiltonian and changing the degeneracy of the ordered states.\cite{Chen} We consider one example in the Appendix, in which the result is the appearance of an intermediate phase between the Coulomb and ordered phases.

\section{Mapping to \kagome\ bosons}
\label{SecMapping}

We now describe the first stage of our derivation of the continuum theory for the transition, in which we map from the statistics of classical dimers to a quantum model. We do so by using the standard mapping between classical statistical mechanics in $d$ dimensions and quantum mechanics in $d-1$ dimensions, in which one spatial dimension of the classical problem is interpreted as the (imaginary) time direction for the quantum problem.

\subsection{Definition of mapping}
\label{SecDefineMapping}

In defining the mapping, we have the freedom to choose the time direction for the quantum problem, and we do so in a way that does not distinguish between the ordering patterns in the low-temperature phase. While one of the cubic axes might seem to be a natural choice, taking this would necessarily distinguish those cases where $\langle \mv \rangle$ is parallel to the time direction from those where it is perpendicular. Instead, we choose the $[111]$ direction and define the quantum imaginary time as $\tau = \sum _\mu r_\mu$.

The mapping follows the standard procedure of using a transfer matrix to connect the degrees of freedom in one layer of the system to those in the next, followed by interpretation of the transfer matrix as the exponential of a quantum Hamiltonian. We first divide the links of the cubic lattice into layers by the imaginary-time coordinates of their midpoints. Each layer is treated as a time slice and the rows and columns of the transfer matrix $\T _1$ are labeled by the configurations of two adjacent layers. The configurations of a given $(111)$ plane are mapped onto the basis states of a quantum Hilbert space by simply identifying the presence (or absence) of a dimer with the presence (absence) of a boson.

\begin{figure}
\putinscaledfigure{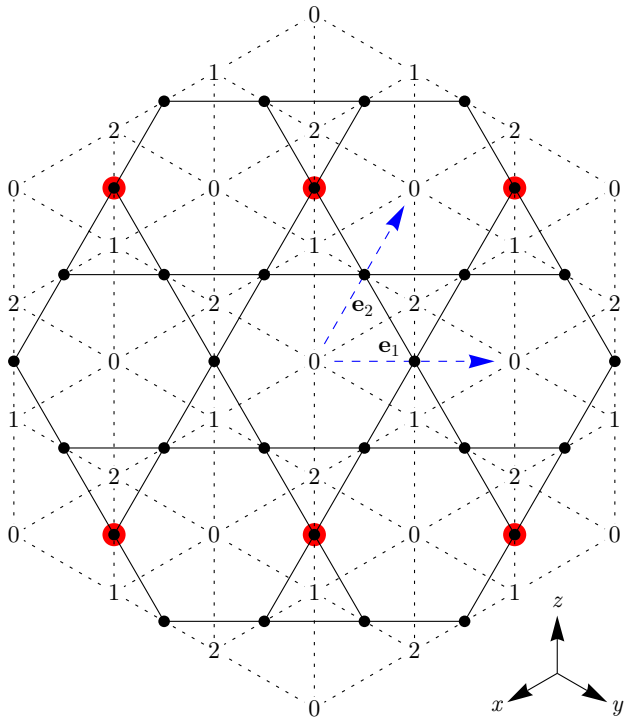}
\caption{\label{FigCubic111Projection}(color online) Projection of the cubic lattice onto a $(111)$ plane, with the \kagome\ lattice superimposed. The cubic sites are shown by the numbers $0$, $1$ and $2$, giving the quantum imaginary-time coordinate $\tau \bmod 3$ (see main text). Points with solid lines show the sites of the \kagome\ lattice, situated at the centres of the cubic bonds (dashed lines) between sites with $\tau \bmod 3 = 1$ and $2$; they therefore lie in planes with $\tau \bmod 3 = \frac{3}{2}$. The larger red circles shown superimposed on some of the \kagome\ sites illustrate the occupied sites in one of the six degenerate ordering patterns, corresponding to the six ordered states of the cubic dimers. The elementary unit vectors of the \kagome\ lattice, $\ev _1$ and $\ev _2$, are shown with dashed blue arrows. The coordinate axes shown in the bottom-right of the figure are the projection of the cubic axes onto the $(111)$ plane.}
\end{figure}
In order for the resulting quantum model to describe lattice bosons, we must ensure two features. First, we require the sites of the lattice in one time slice to correspond to those in the next, and second, we require conservation of particle number. As far as the first requirement is concerned, Figure~\ref{FigCubic111Projection} shows that the midpoints of cubic links with a given value of $\tau$ form a \kagome\ lattice, and that the lattices formed by adjacent layers are displaced with respect to each other. Our first requirement can nonetheless be satisfied by taking a product of (any multiple of) three elementary transfer matrices $\T _1$ to define a single time step. As illustrated in Figure~\ref{FigCubic111Projection}, planes separated by three units in the time direction coincide.\footnote{An alternative approach would be to define a transfer matrix consisting of $\T _1$ followed by a translation (or other transformation in the plane) to return the sites to their original locations. Such a choice would necessarily reduce the symmetry of the quantum Hamiltonian that results.} (Each of the three elementary matrices has a different form because of the relative displacements between successive \kagome\ layers, but for simplicity we denote them all without distinction by $\T _1$.)

We also require conservation of particle number, meaning that the number of bosons in a given time slice should be constrained to equal the number in the following. The close-packing constraint in \refeq{EqClosePacking} implies that
\beq{EqClosePacking2}
\sum _{\rv \in \tau} \sum _\mu d_\mu(\rv) = \sum _{\rv \in (\tau - 1)} \left [ 1 - \sum _\mu d_\mu(\rv) \right]\punc{,}
\eeq
where $\rv \in \tau$ indicates a sum over all cubic sites in imaginary-time slice $\tau$. This equation states that if the \kagome\ plane at $\tau - \frac{1}{2}$ has $n$ bosons and a total of $A$ sites, then the plane at $\tau + \frac{1}{2}$ will have $\frac{A}{3} - n$ bosons. To define a transfer matrix that conserves particle number, we must therefore take a product of an even number of elementary transfer matrices $\T _1$.

To satisfy the two requirements, we must therefore define the full transfer matrix $\T = \T _1^{\delta\tau}$ with $\delta\tau$ an integer divisible by both $3$ and $2$, and so we take $\delta\tau = 6$. The transfer matrix $\T$ therefore has rows and columns labeled by the configurations of two planes separated by $\delta\tau = 6$, and its elements give the statistical weights for those configurations, summed over all possible configurations of the intermediate planes.

The partition function for the classical problem is given by $\Z = \Tr \T^{L/\delta \tau}$, where $L$ is the length of the system in the $[111]$ direction and periodic boundary conditions are assumed in this direction. The effective quantum Hamiltonian $\quH$ is defined by $\T = \ee^{-\quH \delta\tau}$, so that $\Z$ is given by the quantum partition function at inverse temperature $\beta \propto L$. The classical thermodynamic limit is therefore given by the quantum zero-temperature limit, $\beta \rightarrow \infty$, and we will always work in this limit.

\subsection{Quantum Hamiltonian}
\label{SecQuantumHamiltonian}

For a finite lattice, it is in principle possible to find the transfer matrix $\T$ exactly, by considering all allowed configurations of two planes separated by $\delta \tau = 6$ and summing the Boltzmann weights of all possible arrangements of the intermediate planes. The quantum Hamiltonian $\quH$ can then be found by taking the (matrix) logarithm of $\T$. For even fairly small lattices, however, the number of configurations is large, making this a computationally difficult problem, and we have not attempted to find $\T$ or $\quH$ exactly.

Since we are interested in the long-wavelength properties of the model near the transition, we will instead use general considerations such as symmetry to determine the form of the Hamiltonian. As we have noted above, the Hamiltonian describes the dynamics of bosons on a \kagome\ lattice, with conserved particle number. The hard-core nature of the dimers implies that the bosons have a similar hard-core constraint, restricting occupation numbers to zero or one on any site of the lattice.

Further, the close-packing constraint in \refeq{EqClosePacking}, which ensures particle-number conservation, also implies that any triangle of the \kagome\ lattice, of either orientation, can be occupied by at most one boson. This implies a nearest-neighbour repulsion of infinite magnitude.

We define the number operator for \kagome\ site $i$ as $n_i$, and (hard-core) bosonic creation and annihilation operators $b^\dagger _i$ and $b^{\phantom{\dagger}}_i$. The most general quantum Hamiltonian, with conserved particle number and obeying the close-packing constraint, can then be written in the form
\beqm{EqQuantumHamiltonian}
\quH = - \mu \sum _i n_i + \frac{U}{2} \sum _i n_i(n_i - 1) + U\sum _{\langle i j \rangle} n_i n_j \\{}+ \sum_{i,j} V_{ij} n_i n_j - \sum _{i,j} t_{ij} b^\dagger_i b^{\phantom{\dagger}}_j + \cdots\punc{,}
\end{multline}
where the ellipsis represents other terms, such as three-body interactions and correlated hopping terms,\cite{SpinIce} whose precise form is not important. The hard-core constraints have been represented by interactions of strength $U \rightarrow \infty$, the coefficients $V_{ij}$ describe a further-neighbor repulsion, and $t_{ij}$ is the hopping. While the Hamiltonian conserves particle number, summing over all classical configurations means that all particle-number sectors should be included in the quantum-mechanical trace. The effective quantum problem is therefore defined in the grand-canonical ensemble, with a chemical potential $\mu$ that, like the other coefficients, emerges as an effective parameter.

A generalization of the dimer model\cite{Charrier} to the case where the occupation number of a given link is allowed to take on values other than $0$ or $1$ would result in a similar quantum Hamiltonian, but with a correspondingly expanded on-site Hilbert space. An alternative generalization of the dimer model allows for `monomers', where the close-packing constraint in \refeq{EqClosePacking} can be violated to permit a site touched by zero or multiple dimers (with a finite energy cost). Such a modification breaks conservation of particle number and adds to $\quH$ terms such as $\sum _i (J_i b_i^\dagger + J_i^* b_i)$ which eliminate the phase-rotation symmetry.

Note that, as is generally the case for effective quantum Hamiltonians used to describe classical partition functions, it is not necessary for $\quH$ to be hermitian.\cite{Hatano} For example, the hopping coefficients are in general not symmetric, $t_{ij} \neq t_{ji}^*$, following from the fact that choosing a particular $(111)$ plane to define the quantum problem breaks time-reversal symmetry. As we have discussed previously in regard to a related problem in spin ice,\cite{SpinIce} the nonhermitian terms are crucial for reproducing the correct spatial dependence of the long-range correlation functions.

\subsubsection*{Locality}

For the analysis that follows, an important condition on the Hamiltonian $\quH$ is that it should be local, at least when projected into a suitable subspace of low-energy states. While the classical configuration energy $\clH$ and the close-packing constraint in \refeq{EqClosePacking} are local, this does not necessarily imply the same for the transfer matrix or quantum Hamiltonian. We have no general proof that the locality condition is satisfied, and it is in fact possible, by considering states with sufficiently high energy, to construct configurations on which the effect of $\T$ is to cause hopping over arbitrarily long distances.

We argue, however, that locality is satisfied in the region of interest, at low energy near the transition. In this region, low-energy configurations of the original cubic dimer problem can be described in terms of ordered regions separated by two-dimensional domain walls, which cost an energy proportional to their surface area. The intersections of these with a given time slice give the one-dimensional domain walls of the quantum problem, separating different density-wave orderings of the bosons. Consider, in the quantum picture, a time step in which a domain wall moves by a large distance, so that one of the two neighboring domains grows by an area $\delta A$. In the classical partition function, this corresponds to a configuration where a domain wall has a section of area $\delta A$ running parallel to the $(111)$ plane, in between the two consecutive time slices. Such a configuration has an energy cost that grows linearly with $\delta A$, and hence has an exponentially suppressed contribution to the transfer matrix.

Further confirmation of the applicability of a local Hamiltonian comes from the analysis of the Coulomb phase in Section~\ref{SecCoulombPhase}, which reproduces the correct power-law form of the long-range correlation functions [see \refeq{EqCoulombCorrelations}] on the assumption that the low-energy excitations in the superfluid phase are phonons with a linear dispersion. A definitive answer to the question of locality could of course be found by computing the quantum Hamiltonian exactly on a sufficiently large lattice.

\subsection{Phase structure}
\label{SecHamPhases}

\begin{figure}
\putinscaledfigure{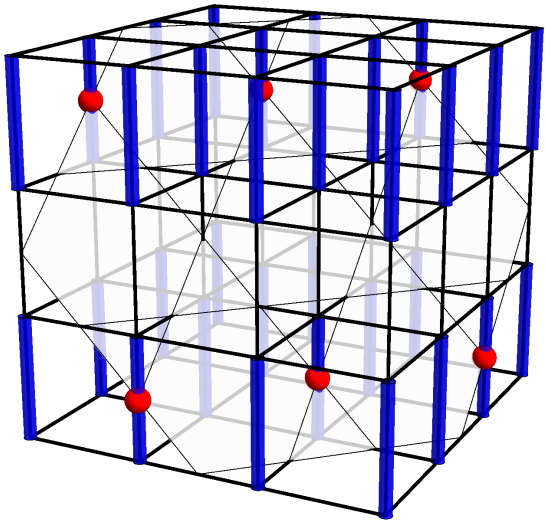}
\caption{\label{FigCubicKagome3D}(color online) Part of the cubic lattice showing one of the ordered states of the dimers (in blue). A $(111)$ plane is superimposed on the crystal structure, cutting diagonally through the cube. Such planes comprise the `time slices' on which the quantum Hilbert space is defined. The sites of the two-dimensional problem are situated where the $(111)$ plane intersects links of the cubic lattice; these form a \kagome\ lattice. Where a cubic link is occupied by a dimer, the corresponding \kagome\ site is occupied by a boson (shown with red spheres).}
\end{figure}
A major advantage of this particular choice of mapping is that the six distinct ordering patterns for the classical dimers map to six ordered states of the quantum problem, related to each other by symmetry. As noted in Section~\ref{SecModel}, the states that minimize the classical configuration energy $\clH$ have all dimers parallel to one of the three cubic axes, and half of the links of this orientation occupied. An example is shown in Figure~\ref{FigCubicKagome3D}, along with the corresponding arrangement of quantum bosons.

Cubic links with the same orientation map onto the same \kagome\ sublattice, and since half of the cubic links of a given orientation are occupied, the same is true of the \kagome\ sites of a given sublattice. As illustrated in Figure~\ref{FigCubicKagome3D}, the classical configurations that minimize $\clH$ map to density-wave states of the bosons, at filling (bosons per site) of $\frac{1}{6}$. The three states with $\mv = +\deltav _{x,y,z}$ are related by a rotation of the \kagome\ plane, while $\mv = \pm \deltav _z$ are related by a translation.

As the (classical) temperature is raised from zero, thermal fluctuations in the dimer configuration will occur, with those of lowest energy being single flipped plaquettes. In terms of bosons, these correspond to quantum fluctuations away from the perfectly ordered density-wave patterns and occur once the hopping coefficients $t_{ij}$ in \refeq{EqQuantumHamiltonian} become nonzero. For sufficiently small hopping $t_{ij}$ relative to the further-neighbor repulsion $V_{ij}$, the ground state remains ordered. We identify this phase, where (connected) correlation functions are short-ranged and the lattice symmetry is broken, as a Mott insulator of bosons with density-wave order.

When the classical temperature is raised beyond the critical value $T\sub{C}$, the dimer order is lost. In the resulting high-temperature Coulomb phase, the full lattice symmetry is restored and each cubic link has an equal average occupation of $\frac{1}{6}$. In the quantum model, the transition corresponds to a loss of density-wave order at a critical hopping $t_{ij}$, and restored lattice symmetry implies a uniform quantum ground state with average particle number $\frac{1}{6}$ on each site. (Note that, since there are three sites in the \kagome\ unit cell, this filling corresponds to $\frac{1}{2}$ per unit cell. The hard-core repulsion between nearest-neighbor sites also means that filling $\frac{1}{6}$ is equal to half of the maximum possible filling.)

This uniform ground state can be identified with the superfluid simply by noting that, at fractional filling and with neither quenched disorder nor spatial symmetry breaking, the only possible phase for quantum bosons at zero temperature is a condensate. In Section~\ref{SecCoulombPhase}, we will show that the power-law correlations within the Coulomb phase are correctly reproduced by the phase mode of the condensate, providing further justification for the identification of these phases.

As an aside, we consider the insight into this equivalence that comes from the phenomenon of off-diagonal long-range order (ODLRO).\cite{SpinIce} The existence of a nonzero superfluid order parameter implies that, in the condensed phase, the quantum expectation value $\langle b^\dagger_i b^{\phantom{\dagger}}_j \rangle$ approaches a nonzero constant in the limit of large separation of the points $i$ and $j$. By contrast, no such ODLRO exists in the Mott insulator and the limiting value is zero.

The quantum expectation value is defined by
\beq{ODLRO}
\langle b^\dagger_i b^{\phantom{\dagger}}_j \rangle = \frac{\Z_{ij}}{\Z} = \frac{1}{\Z} \Tr \left( \T^{L/\delta\tau} b^\dagger_i b^{\phantom{\dagger}}_j \right)\punc{.}
\eeq
The quantity $\Z_{i,j}$ can be understood as the sum over histories of the quantum problem, with a particle creation event at site $i$ and a particle annihilation at site $j$, on the same (arbitrary) time slice. Returning to the language of the classical statistical problem, these events at which particle conservation is broken become, according to the arguments following \refeq{EqClosePacking2}, points in three-dimensional space where the close-packing constraint is violated. One should therefore understand $\Z_{i,j}$ as the partition function of the dimer model calculated in the presence of two test monomers at positions $i$ and $j$ on the same (arbitrary) time slice. (Strictly, one is an empty site, while the other is a site where two dimers meet at a site.)

In the low-temperature phase, such monomers disrupt the ordering pattern and cost energy proportional to the linear separation between the two sites. By contrast, in the Coulomb phase monomers are deconfined, separating them to infinity costs a finite energy,\cite{Huse} and $\Z_{i,j}$ approaches a nonzero limit for large separation. The ODLRO in the quantum superfluid is therefore equivalent to the deconfinement of monomers; this is consistent with our identification of the Coulomb and superfluid phases.

\subsection{Symmetries of the Hamiltonian}
\label{SecHamSymmetries}

As we will show in Sections~\ref{SecCoulombPhase} and \ref{SecTransition}, an understanding of the behavior both deep in the Coulomb phase and near the ordering transition depends on an analysis of the symmetries of the quantum model. We treat here the case where the classical configuration energy $\clH$ has full cubic symmetry. Chen et al.\cite{Chen}\ have studied the effects of various modifications that reduce this symmetry, and we consider one example in the Appendix.

First, since the quantum Hamiltonian $\quH$ describes a model with conserved particle number, it has a \U1 symmetry under global phase rotations of the bosonic creation and annihilation operators: $b_i \rightarrow b_i \ee^{\ii \vartheta}$ and $b_i^\dagger \rightarrow b_i^\dagger \ee^{-\ii\vartheta}$. This symmetry is spontaneously broken in the superfluid phase, where $b_i$ has a nonzero expectation value.

\begin{figure}
\putinscaledfigure{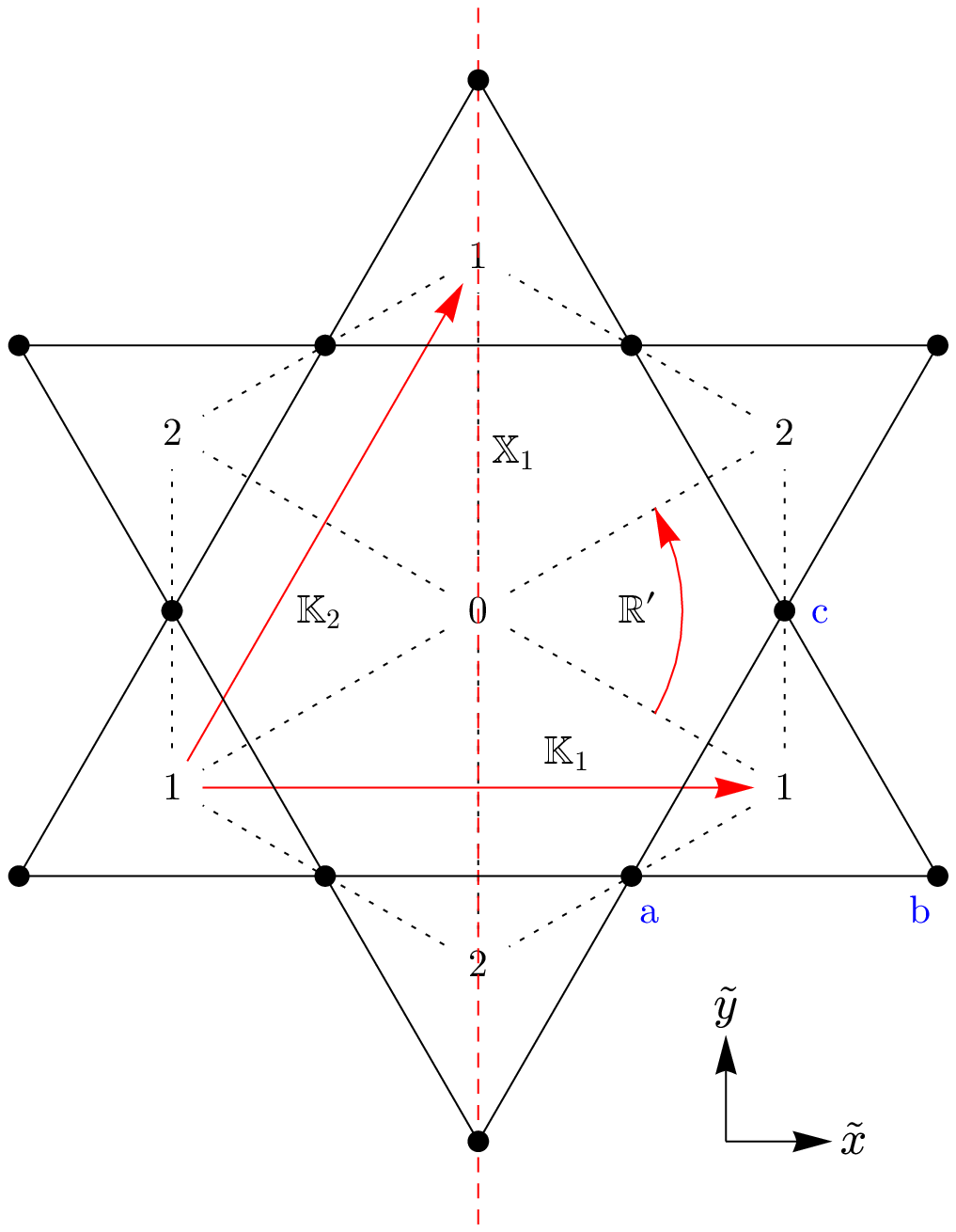}
\caption{\label{FigKagomeSymmetries}(color online) Part of the \kagome\ lattice, shown with points joined by solid lines, with the projection of the cubic lattice superimposed, as in Figure~\ref{FigCubic111Projection}. Four symmetry operations, $\KagomeTrans _1$, $\KagomeTrans _2$, $\Rot '$, and $\RfX _1$, are illustrated in red. The two primitive translations $\KagomeTrans _1$ and $\KagomeTrans _2$ are shown with straight arrows, while $\Rot '$, a rotation by $60\degrees$ about the center of a \kagome\ hexagon, is shown with a curved arrow. The dashed red vertical line shows the line of reflection for $\RfX _1$. The three sites of the triangle at the bottom right are labeled $\sublattice{a}$, $\sublattice{b}$, and $\sublattice{c}$ to denote the three \kagome\ sublattices.}
\end{figure}
Besides this internal symmetry, there are also spatial symmetries inherited from those of the classical dimer model, but modified in important ways by the particular choice of the imaginary-time direction. These symmetries can be constructed from combinations of three primitive symmetry operations, a translation $\KagomeTrans _1$, a rotation $\Rot$, and a reflection $\RfX _1$. These operations are illustrated in Figure~\ref{FigKagomeSymmetries}, along with the translation $\KagomeTrans _2 = \Rot \KagomeTrans _1 \Rot^{-1}$.

We parametrize the positions of lattice sites in the \kagome\ planes by orthogonal coordinates $\kagx$ and $\kagy$:
\beq{EqKagomeCoordinates}
\begin{split}
\kagx &= \sqrt{\frac{3}{2}} (r_y - r_x)\\
\kagy &= \sqrt{2} \left(r_z - \frac{1}{2}r_x - \frac{1}{2}r_y\right)\punc{,}
\end{split}
\eeq
where $r _\mu$ are coordinates referred to the cubic axes, which take integer values at the cubic lattice sites.

The two operators $\KagomeTrans _1$ and $\KagomeTrans _2$ perform translations by the elementary unit vectors $\ev _1$ and $\ev _2$ of the \kagome\ lattice, 
transforming the coordinates $\kagx$ and $\kagy$ according to
\begin{align}
\label{EqKagomeTrans}
\begin{pmatrix}
\kagx \\ \kagy
\end{pmatrix}
&\xrightarrow{\KagomeTrans _1}
\begin{pmatrix}
\kagx \\ \kagy
\end{pmatrix}
+
\sqrt{6}
\begin{pmatrix}
1 \\ 0
\end{pmatrix}
\\
\begin{pmatrix}
\kagx \\ \kagy
\end{pmatrix}
&\xrightarrow{\KagomeTrans _2}
\begin{pmatrix}
\kagx \\ \kagy
\end{pmatrix}
+
\sqrt{6}
\begin{pmatrix}
1/2 \\ \sqrt{3}/2
\end{pmatrix}
\end{align}
They can be expressed in terms of pairs of translation operators $\CubicTrans _\mu$ for the cubic lattice, chosen so that the imaginary time coordinate $\tau$ is unchanged; for example, $\KagomeTrans _1 = \CubicTrans _x^{-1} \CubicTrans _y$. These transformations map the three \kagome\ sublattices to themselves.

We define $\Rot '$ as a rotation by $60\degrees$ about the center of a \kagome\ hexagon, or, in terms of the cubic lattice, about a $[111]$ axis passing though a cubic site with $\tau \bmod 3 = 0$ (such as the one at the center of Figure~\ref{FigKagomeSymmetries}). This maps the \kagome\ lattice to itself, but is not a symmetry of the cubic lattice, since, as can be seen in Figure~\ref{FigKagomeSymmetries}, it exchanges cubic sites with $\tau \bmod 3 = 1$ and $\tau \bmod 3 = 2$. We therefore define the operation $\Rot$ consisting of an improper rotation by $60 \degrees$ through this axis, which is a symmetry of the cubic lattice. In terms of the bosons, $\Rot$ consists of the rotation $\Rot '$ followed by a time-reversal operation $\tau \rightarrow 3 - \tau$, and it commutes with quantum Hamiltonian $\quH$. (A similar time-reversed symmetry operation was found to apply in an effective quantum description of spin ice.\cite{SpinIce} The absence of time reversal as an independent symmetry reflects the fact, noted in Section~\ref{SecQuantumHamiltonian}, that $\quH$ is not hermitian.) The rotation $\Rot$ permutes the three \kagome\ sublattices cyclically.

The remaining primitive symmetry operation of the \kagome\ model is the reflection $\RfX _1$ through a line running perpendicular to the unit vector $\ev _1$. It transforms $\kagx \rightarrow -\kagx$, and exchanges two of the sublattices (labeled $\sublattice{a}$ and $\sublattice{b}$) while leaving the third unchanged.

Besides these symmetries of the effective quantum Hamiltonian, there are further symmetries of the original classical model that are broken by the explicit choice of the $[111]$ direction as imaginary time. These include reflections in the cubic $(100)$, $(010)$, and $(001)$ planes, which we denote $\RfI _x$, $\RfI _y$, and $\RfI _z$ respectively, and which relate different equivalent choices of the imaginary-time direction. They cannot be written as operations on the quantum Hilbert space, but in terms of the continuum space-time action to be derived below, they are simply reflections.

As an aside, one can also consider the operator representing translation in the time direction by $3$ steps. As noted in Section~\ref{SecDefineMapping}, the full transfer matrix $\T = \T _1^6$ connects two $(111)$ planes separated by $6$ steps in the imaginary-time direction, and is the simplest choice consistent with conservation of particle number. We can nonetheless consider the operator $\T^{1/2} = \T _1^3$ representing `imaginary-time evolution' by three steps, which maps the \kagome\ lattice to itself, and clearly commutes with $\T$ and hence with the quantum Hamiltonian $\quH$.

Unlike the other symmetries of $\quH$, $\T^{1/2}$ cannot be written as a permutation matrix in the occupation-number basis, and as for the full transfer matrix, we have not attempted to find its precise form. For our purposes, it is sufficient to note its effect on the density: as follows from the observations of Section~\ref{SecDefineMapping}, if a given \kagome\ plane has a density of $\rho$ bosons per site, then the plane $3$ steps later has density $\frac{1}{3} - \rho$. Following our assumption of the locality of $\T$, this implies that, in the coarse-grained limit, $\T^{1/2}$ simply changes the sign of local density fluctuations. The microscopic Hamiltonian is not invariant under a particle--hole transformation, and this is therefore an emergent symmetry of the long-wavelength limit.

\section{Continuum theory for Coulomb phase}
\label{SecCoulombPhase}

As noted in Section~\ref{SecModel}, above the critical temperature $T\sub{C}$, the classical dimer model exhibits a Coulomb phase, in which there is no ordering, but long-range correlation functions have power-law forms and strong spatial dependence. This behavior can be understood in terms of a coarse-grained picture, in which the long-wavelength degrees of freedom are described by a solenoidal field.\cite{Huse} Deep within the Coulomb phase, this approach predicts the dipolar form for the dimer-dimer correlation function given in \refeq{EqDipolarCorrelations}.

In this section, we will show that the long-distance behavior of the correlation functions can also be obtained from the effective quantum model derived in Section~\ref{SecMapping}. The power-law behavior of the correlation functions follows immediately from the presence of a Goldstone phase mode in the superfluid, while the precise spatial dependence of the dipolar correlations can be reproduced by taking into account the symmetries of the effective quantum Hamiltonian. We have applied a similar analysis to a related model of spin ice.\cite{SpinIce}

\subsection{\Kagome\ continuum action}
\label{SecKagomeContinuum}

To describe the long-distance properties of the superfluid phase of the quantum model, we pass from the microscopic description to a continuum action. This action is written in terms of bosonic fields $\Psi$ corresponding to the hard-core boson operator $b$ in the limit where the spatial coordinates $\kagx$ and $\kagy$ and the imaginary time $\tau$ are taken as continuous. To preserve the important effects of the \kagome\ lattice structure, we define three ({\it c}-number) fields $\Psi _\sigma$ corresponding to the three \kagome\ sublattices, $\sigma \in \{\sublattice{a},\sublattice{b},\sublattice{c}\}$ (illustrated in Figure~\ref{FigKagomeSymmetries}).

The continuum action $\Act$ will contain all powers of the fields and their derivatives consistent with the symmetries of the hard-core boson problem, described in Section~\ref{SecHamSymmetries}. We must therefore determine the effects of these symmetries on the fields and the coordinates $\kagx$, $\kagy$, and $\tau$.

Firstly, the symmetry under uniform \U1 phase rotations of the boson operators leads to the same condition on the fields $\Psi _\sigma$ and hence restricts the terms in $\Act$ to be those which are invariant under such phase rotations. The translation operators $\KagomeTrans _1$ and $\KagomeTrans _2$ do not affect the sublattices and simply lead to translations of the coordinates $\kagx$ and $\kagy$.

Next, consider the operator $\Rot$, which consists of a rotation by $60\degrees$ followed by a time-reversal operation. The rotation permutes the three sublattices cyclically and also acts on the spatial coordinates, while time reversal corresponds to complex conjugation of the field operators\cite{SpinIce} $\Psi _\sigma^{\phantom{*}} \rightarrow \Psi _\sigma^*$. Finally, the reflection $\RfX _1$ exchanges sublattices $\sublattice{a}$ and $\sublattice{b}$ while also reflecting the coordinate $\kagy \rightarrow -\kagy$.

These transformations can be written in a simpler form by defining the derivative operators $\kpar _\pm = \pm \frac{\sqrt{3}}{2} \partial _{\kagx} - \frac{1}{2}\partial_{\kagy}$, the vectors $\kparv$ and $\Psiv$,
\beq{EqVectors}
\kparv =
\begin{pmatrix}
\kpar _-\\
\kpar _+\\
\partial _{\kagy}
\end{pmatrix}\qquad\text{and}\qquad
\Psiv =
\begin{pmatrix}
\Psi _{\sublattice{a}}\\
\Psi _{\sublattice{b}}\\
\Psi _{\sublattice{c}}
\end{pmatrix}\punc{,}
\eeq
and the matrices $\mathbf{R}$ and $\mathbf{X}_1$,
\beq{EqMatrices}
\mathbf{R} = \begin{pmatrix}
0 & 1 & 0\\
0 & 0 & 1\\
1 & 0 & 0
\end{pmatrix}
\qquad\text{and}\qquad
\mathbf{X}_1 = \begin{pmatrix}
0 & 1 & 0\\
1 & 0 & 0\\
0 & 0 & 1
\end{pmatrix}\punc{.}
\eeq
The effects of the symmetry operations $\Rot$ and $\RfX _1$ on $\kparv$, $\Psiv$, and the time derivative $\partial _\tau$ are summarized in Table~\ref{TabSymmetries}.
\begin{table}
\caption{\label{TabSymmetries}Effects of the symmetry operators $\Rot$ and $\RfX _1$ on the continuum fields and derivative operators, expressed in terms of $\kparv$ and $\Psiv$ [\refeq{EqVectors}], $\mathbf{R}$ and $\mathbf{X}_1$ [\refeq{EqMatrices}], $\phi$ [\refeq{EqActionPhi}], and $\nv$ [\refeq{EqDensityField}], as well as the cubic lattice position vector $\rv$, derivative $\parv$, and dimer density field $\dv$, discussed in Section~\ref{SecCubicContinuum}.}
\begin{ruledtabular}
\begin{tabular}{ccc}
\phantom{$\Rot$} & $\Rot$ & $\RfX _1$ \\
\hline
$\kparv$ & $-\mathbf{R}\kparv$ & $\mathbf{X}_1\kparv$ \\
$\partial _\tau$ & $-\partial _\tau$ & $\partial _\tau$\\
\hline
$\Psiv$ & $\mathbf{R}\Psiv^*$ & $\mathbf{X}_1\Psiv$ \\
$\phi$ & $-\phi$ & $\phi$ \\
$\nv$ & $\mathbf{R} \nv$ & $\mathbf{X}_1 \nv$ \\
\hline
$\rv$ & $-\mathbf{R}\rv$ & $\mathbf{X}_1 \rv$\\
$\parv$ & $-\mathbf{R}\parv$ & $\mathbf{X}_1\parv$\\
$\dv$ & $\mathbf{R} \dv$ & $\mathbf{X}_1 \dv$
\end{tabular}
\end{ruledtabular}
\end{table}

These symmetry considerations allow a continuum action to be written in terms of the field $\Psiv$, which should contain all terms that are invariant under the action of the full symmetry group.

In the Coulomb phase of the classical model, the quantum bosons condense, so that the field $\Psiv$ acquires a nonzero expectation value. This phase breaks no spatial symmetries, and so $\langle \Psi _\sigma \rangle$ is equal on the three sublattices $\sigma$. Away from the transition, the long-wavelength properties are dominated by the gapless Goldstone mode describing uniform rotations of the phase, and corresponding to the broken \U1 symmetry. Writing $\Psi _\sigma \sim \ee^{\ii \phi}$, the effective action can be expressed in terms of the field $\phi$, whose symmetry properties are shown in Table~\ref{TabSymmetries}.

The effective phase-only action $\Act _\phi$ can be written in the form
\beq{EqActionPhi}
\Act _\phi = \int \dd \kagx \, \dd \kagy \, \dd \tau \left [ - \phi (\kparv^2 + \partial _\tau^2) \phi + \cdots \right]\punc{,}
\eeq
where the ellipsis denotes terms with higher powers of $\phi$ or higher derivatives. (All terms with a single derivative can be rewritten as total derivatives and so vanish on integration.) This continuum action is explicitly space-time symmetric; the relative coefficients of the spatial and temporal derivatives are required to be equal [in the appropriate units, chosen in \refeq{EqKagomeCoordinates}] by the full cubic symmetry of the original model (or equivalently by symmetry under the inversion operators $\RfI _\mu$).

To evaluate the correlation functions of the dimer occupation numbers, we must relate these quantities to the continuum field $\phi$. First, consider the boson density, which we represent by the field $n_\sigma$ giving the local density measured relative to the average filling of $\frac{1}{6}$. The symmetry properties of the vector $\nv$ are included in Table~\ref{TabSymmetries}; note that it is invariant under time reversal. These properties are sufficient to identify the density operator (to leading order) as
\beq{EqDensityField}
n_\sigma \sim \kpar _\sigma \phi + \partial _\tau \phi\punc{,}
\eeq
where the relative coefficient is again fixed using cubic symmetry. (The first term is allowed only because of the absence of time-reversal symmetry, a consequence of the nonhermitian nature of $\quH$ discussed in Section~\ref{SecQuantumHamiltonian}.)

\subsection{Cubic lattice}
\label{SecCubicContinuum}

The symmetry operations $\Rot$ and $\RfX _1$ have so far been treated as acting within the \kagome\ planes, but they are also symmetries of the full cubic lattice. Their action (in the continuum limit) on the three-dimensional position vector $\rv$ and derivative operator $\parv$ are shown in Table~\ref{TabSymmetries}, along with the behavior of the field $\dv$, which is defined as the continuum limit of the dimer occupation number, with the average occupation of $\frac{1}{6}$ subtracted.

To determine the correlation functions of the dimer field $\dv$, we must relate it to $\phi$. It would be consistent with the symmetries listed in Table~\ref{TabSymmetries} to identify $d_\mu$ with the combination $\partial _\mu \phi$. This is, however, incorrect, as can most easily be seen by making use of the cubic reflections $\RfI _\mu$, defined in Section~\ref{SecHamSymmetries}: $\partial _\mu \phi$ changes sign under $\RfI _\mu$, whereas $d _\mu$ does not. To find the correct relationship between these fields, consider the effect of the cubic reflections on the microscopic dimer degrees of freedom, such as $d_x (\zerov)$, which gives the occupation number for the link between the sites $\zerov$ and $\deltav_x$. Under $\RfI _x$, this maps to the link between $-\deltav_x$ and $\zerov$, described by $d _\mu (-\deltav _x)$; in general, the microscopic variable $d _\nu (\rv)$ maps to $d _\nu (\RfI _\mu \rv - \delta_{\mu \nu} \deltav _\mu)$ under $\RfI _\mu$. Using $\eta _\rv$, equal to $\pm 1$ on the two sublattices, we can therefore construct the combination $\eta _\rv d _\mu(\rv)$, which, after coarse-graining, changes sign under $\RfI _\mu$ (since $\eta _\rv = -\eta _{\RfI_\mu \rv - \deltav _\mu}$ for any $\mu$). The symmetries in Table~\ref{TabSymmetries} are insufficient in this case because they all map from one quantum plane to another; these are separated by multiples of $\delta \tau = 6$, and so $\eta _\rv = (-1)^\tau$ is unchanged. One can instead use the operator $\T^{1/2}$, defined at the end of Section~\ref{SecHamSymmetries}. Particle-number conservation between adjacent planes with $\tau \bmod 6 = \frac{3}{2}$ leads to \U1 symmetry under rotations of the phase $\phi$. This symmetry therefore acts with the opposite sign in the planes with $\tau \bmod 6 = \frac{3}{2} + 3$, leading to the factor of $\eta _\rv = (-1)^\tau$.

We can therefore identify $\eta _\rv d_\mu \sim \partial _\mu \phi$, which, together with the action given in \refeq{EqActionPhi}, allows the Coulomb-phase dimer-dimer correlation function to be found. The (imaginary-time ordered) propagator for the field $\phi$ is simply $1/|\kv|^2$, leading\cite{SpinIce} to real-space correlations with a dipolar form,
\beq{EqCoulombCorrelations}
\langle d_\mu(\rv) d_\nu(\zerov) \rangle \sim \eta _\rv \frac{3 r_\mu r_\nu - |\rv|^2 \delta _{\mu\nu}}{{|\rv|}^5}\punc{.}
\eeq
This form for the correlators was predicted by Huse et al.,\cite{Huse} by considering the continuum limit of a coarse-grained action for the dimer degrees of freedom.

\section{Continuum theory of transition}
\label{SecTransition}

The mapping described in Section~\ref{SecMapping} relates the thermal transition between a dimer crystal and a Coulomb phase to the quantum phase transition from a Mott insulator with density-wave order to a superfluid. A continuum theory to describe the phase transition in the dimer model can therefore be found by considering this equivalent quantum transition. As noted in Section~\ref{SecIntroduction}, the presence of two incompatible order parameters makes the standard LGW approach insufficient, and instead, the critical theory can be found using a mapping to dual vortex fields.

\subsection{Duality mapping}

This duality mapping has been described in detail by Balents et al.,\cite{Balents} and we will simply sketch a derivation. (Note that Sengupta et al.\cite{Sengupta}\ found the critical theory for a filling factor of $f = \frac{1}{3}$ on the \kagome\ lattice.) The starting point is a current-loop representation of the quantum boson problem,\cite{Wallin} where the degrees of freedom are the currents $\Jv$ defined on the links of the space-time lattice, obeying the continuity equation $\divv \Jv = 0$ (where $\divv$ represents the lattice divergence). The essence of the duality mapping is a transformation from $\Jv$ to a gauge field $\Av$ on the links of the dual lattice, according to $\Jv = \curl \Av$.

It should be noted that mapping to an action written in terms of currents, which can be done using a `Villain representation' for the hopping terms,\cite{Balents,Wallin} involves eliminating all but the nearest-neighbor hopping. (This can be performed explicitly by introducing extra auxiliary fields analogous to $\Jv$ to describe further-neighbor processes, before integrating these out to give renormalized couplings for the currents $\Jv$.) Reflection symmetries such as $\RfX_1$ ensure that the nearest-neighbor hopping coefficients $t_{ij}$ are symmetric, and so the nonhermitian nature of $\quH$ has no effect on the continuum theory near the transition. (It was similarly found in a related model for spin ice\cite{SpinIce} that the directed hopping terms in the effective quantum Hamiltonian were irrelevant at the transition.)

\begin{figure}
\putinscaledfigure{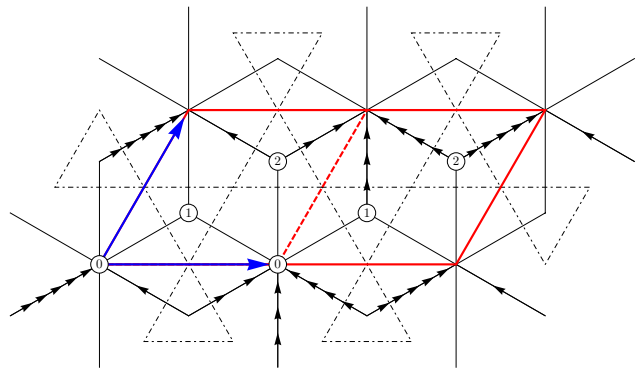}
\caption{\label{FigKagomeDice}(color online) The dice lattice (shown with thin solid lines), dual to the \kagome\ lattice (dashed lines). The unit vectors $\ev _1$ (horizontal) and $\ev _2$, as in Figure~\ref{FigCubicKagome3D}, are shown with blue arrows. The thick red lines show two unit cells of the dice lattice, or one magnetic unit cell. Within this unit cell, the sites of the dice lattice are indicated with black circles containing numbers labeling the three sublattices, $\sigma = 0,1,2$. The background gauge field $\Ab _\ell$ is shown, up to an integer, with black arrows, where each arrowhead represents $\frac{1}{6}$ of a flux unit. The arrangement is chosen so that the curl (defined as the sum of $\Ab _\ell$ going counterclockwise around a loop) is equal to $f = \frac{1}{6}$ for every plaquette. (Moving to the right by $2 \ev _1$, or one magnetic unit cell, increases $\Ab _\ell$ by $1$.)}
\end{figure}

In our approach, the bosons occupy the sites of the \kagome\ lattice, and so the space-time lattice is not cubic as in the original dimer problem, but instead consists of stacked \kagome\ planes. As illustrated in Figure~\ref{FigKagomeDice}, the dual of \kagome\ is the dice lattice,\cite{Vidal,Sengupta,Jiang} and so the gauge field $A _\ell$ is defined on the links $\ell$ of a lattice of stacked dice planes.
The sites of the dice lattice form three sublattices, which we label as $\sigma = 0,1,2$, corresponding to the three distinct plaquettes of \kagome . The positions of sites on the dice lattice are given by the (two-component) vector $\xv = a_1 \ev _1 + a_2 \ev _2 + \sigma \zetav$, where $a_1$ and $a_2$ are integers and $\zetav = \frac{1}{3}(\ev _1 + \ev _2)$ is the displacement between the $0$ and $1$ sublattices.

The currents $\Jv$, and hence the gauge field $\Av$, take integer values (due to the discrete nature of the quantum bosons), and so the model is `frustrated', in the sense that there are many space-time configurations that give nearly equal contributions to the action. (This fact is a straightforward consequence of the fractional boson occupation number; many arrangements of bosons with filling $f = \frac{1}{6}$ have similar interaction energies.) It is more convenient to describe this frustration by introducing `matter fields' $\psi _i$ on the sites $i$ of the dual lattice, and promote $\Av$ to a continuous-valued field. (This step can be performed explicitly by using the Poisson summation formula.\cite{Balents}) The frustration on $\Av$ is then lifted, and we shift it by a (position-dependent) constant $\Avb$ to make clear that $\psi$ now carries the frustration.

The dual theory has a gauge invariance resulting from the definition of $A _\ell$, and can be written in terms of the gauge field $A _\ell$ and matter fields $\psi _i$,
\beqm{EqDualAction}
\Act\sub{dual} = \kappa \sum_p |\curl \Av|^2 \\- t\sub{v} \sum _\ell \left(\psi^\dagger _i \ee^{2\pi \ii (A_\ell + \Ab _\ell)} \psi\nd_j+ \text{h.c.} \right)
\\+ \sum _i (r|\psi _i|^2 + u |\psi _i|^4) + \cdots\punc{,}
\end{multline}
where $\sum _p$ sums over plaquettes $p$ of the dual lattice; $\sum _\ell$ sums over dual-lattice links $\ell$, which start at site $i$ and end at site $j$; and $\sum _i$ sums over dual-lattice sites. Detailed derivations leading to \refeq{EqDualAction} have been given by Balents et al.\cite{Balents}

The field $\psi _i$ corresponds to a vortex of the original bosons, and the gauge field induces the long-range interactions between vortices. By duality, the average boson density of $f = \frac{1}{6}$ per site affects the vortices as a flux of $f$ per plaquette. This is represented by the background field $\Avb$ obeying $\curl \Avb = f$; we choose the gauge illustrated in Figure~\ref{FigKagomeDice}.

The action $\Act\sub{dual}$ consists of a vortex field $\psi _i$ with a frustrating hopping term, coupled to a dynamical gauge field $A _\ell$. Following Balents et al.,\cite{Balents} our approach will be to neglect temporarily the interaction terms, and to consider the effect of the frustration on the dispersion of a single vortex. The full dual action, including the interactions, can then be rewritten in terms of the eigenstates of the single-vortex Hamiltonian $\Ham _1$. The precise form of these eigenstates will depend on the details of $\Act\sub{dual}$, but we will use symmetry considerations to effectively block-diagonalize $\Ham _1$.

\subsection{Projective symmetry group}

The problem of a single vortex with frustrated hopping is equivalent to the Hofstadter problem for a charged particle moving on a lattice in the presence of a magnetic field. Choosing a specific gauge for the background field $\Ab$ reduces the spatial symmetry of the Hamiltonian, and it is convenient to introduce a so-called `projective symmetry group' (PSG).\cite{Balents,Wen} We first address the effect of the lattice symmetries in the real-space basis, before taking the Fourier transform to momentum states, in terms of which the eigenstates of $\Ham _1$ can be written.

The single-particle Hilbert space is spanned by position states $\ket{\xv}$ (where $\xv$ denotes a lattice site), and $\Ham _1$ is a sum of hopping terms of the form
\beq{EqHopping1}
\ket{\xv + \bv}\bra{\xv} \ee^{\ii \alpha(\xv,\bv)}\punc{,}
\eeq
for displacements $\bv$ linking sites of the lattice.

The PSG associates with a lattice symmetry $\Qop$ (any of the translations, rotations, and reflections defined in Section~\ref{SecHamSymmetries}), which maps the site $\xv$ to $\Qop(\xv)$, a corresponding operator $\Qhat$ that commutes with the Hamiltonian $\Ham _1$. Its action on a state $\ket{\xv}$ is given by a real-space transformation, accompanied by a gauge transformation,
\beq{Qaction}
\Qhat \ket{\xv} = \ee^{\ii \chi_\Qop(\xv)}\ket{\Qop(\xv)}\punc{.}
\eeq
Because the Hamiltonian has a lower symmetry than the lattice, the operators $\Qhat$ do not obey the group multiplication table for the full lattice symmetry group. Instead, they obey it up to phase factors, determined by the functions $\chi_\Qop$.

Applying $\Qhat$ to the hopping term in \refeq{EqHopping1} gives
\beqm{EqQonHopping1}
\Qhat \ket{\xv + \bv}\bra{\xv} \ee^{\ii \alpha(\xv,\bv)} \Qhat^{-1} \\= \ee^{\pm \ii \alpha(\xv,\bv)} \ee^{\ii \chi_\Qop(\xv + \bv)}\ee^{-\ii \chi_\Qop(\xv)}\ket{\Qop(\xv+\bv)}\bra{\Qop(\xv)}\punc{,}
\end{multline}
where $\pm$ is positive (or negative) if $\Qhat$ is an (anti)unitary operator. Since $\Qop$ is a lattice symmetry, there must be a corresponding term in $\Ham _1$ given by
\beq{EqHopping2}
\ket{\Qop(\xv + \bv)}\bra{\Qop(\xv)} \ee^{\ii \alpha\boldsymbol{(}\Qop(\xv),\Qop(\xv+\bv) - \Qop(\xv)\boldsymbol{)}}\punc{.}
\eeq
The phases $\chi_\Qop(\xv)$ should be chosen for all $\xv$ in order to make the expressions in \refeqand{EqQonHopping1}{EqHopping2} equal.

We therefore require
\beq{PSGeqs}
\ee^{\ii [\chi_\Qop(\xv + \bv) - \chi_\Qop(\xv)]} = \ee^{\ii [\alpha\boldsymbol{(}\Qop(\xv),\Qop(\xv+\bv) - \Qop(\xv)\boldsymbol{)} \mp \alpha (\xv,\bv)]}\punc{.}
\eeq
This gives a set of equations for $\chi_\Qop(\xv)$ which must be solved simultaneously. (For the translations $\KagomeTrans_{1,2}$ and rotation $\Rot$, a solution can be found for $\Qhat$ a unitary operator, but for the reflections such as $\RfX_1$, $\Qhat$ must be chosen antiunitary.\cite{Balents})

Using the choice of gauge illustrated in Figure~\ref{FigKagomeDice}, the transformation operators acting on the state $\ket{\xv + \sigma \zetav}$, with $\xv = a_1\ev_1 + a_2\ev_2$, give
\begin{align}
\label{EqChis1}
\hat{K}_1 \ket{\xv + \sigma\zetav} &= (-1)^{a_2}\ket{\xv + \ev_1 + \sigma\zetav}\\
\label{EqChiK2}
\hat{K}_2 \ket{\xv + \sigma\zetav} &=\ket{\xv + \ev_2 + \sigma\zetav}\\
\hat{R}\,\ket{\xv + \sigma\zetav} &=\ee^{\frac{\ii\pi}{6}(3a_2+5)(2a_1+a_2+2\delta_{\sigma 2})}\ket{\Rot(\xv + \sigma\zetav)}\\
\hat{X}_1\ket{\xv + \sigma\zetav} &= (-1)^{\frac{1}{2}a_2(a_2+1)} \ket{\RfX_1 (\xv + \sigma\zetav)}\punc{,}
\label{EqChis2}
\end{align}
where $\Rot (\xv + \sigma\zetav) = (-a_2 - \delta_{\sigma 1} - \delta_{\sigma 2})\ev_1 + (a_1 + a_2 + \delta_{\sigma 2})\ev_2 + \bar{\sigma}\zetav$ and $\RfX_1 (\xv + \sigma\zetav) = -(a_1 + a_2 + \sigma)\ev_1 + a_2\ev_2 + \sigma\zetav$. Note that $\Rot$ exchanges sublattices $1$ and $2$; we have defined $\bar\sigma$ so that $\bar0=0$, $\bar1=2$, and $\bar2=1$.

The operators $\Qhat$ form a group with multiplication laws equal, up to phase factors, to those of the original space group formed by the operators $\Qop$. These phases can be calculated using the transformations in the position basis, giving
\begin{align}
\label{EqCommutation1}
\hat{K}_1 \hat{K}_2 &= -\hat{K}_2 \hat{K}_1\\
\hat{K}_2 \hat{R} &= \ee^{\ii\pi/3} \hat{R}\hat{K}_1\\
\hat{K}_2 \hat{R}^2 &= \hat{K}_1 \hat{R}^2 \hat{K}_1\\
\hat{R}^6 &= 1\\
\hat{K}_1 \hat{X}_1 \hat{K}_1 &= \hat{X}_1\\
\hat{K}_1 \hat{X}_1 \hat{K}_2 &= \hat{K}_2 \hat{X}_1\punc{.}
\label{EqCommutation2}
\end{align}
These commutation properties depend only on the effective magnetic flux and are independent of the choice of gauge.

Using the real-space transformations given in \refeqs{EqChis1}{EqChis2}, the Fourier transform to momentum space can be performed. We define the reciprocal lattice vectors $\ev^*_1$ and $\ev^*_2$ so that $\ev _i \cdot \ev^*_j = 2\pi \delta _{ij}$ and the momentum-space basis by
\beq{EqMomentumBasis}
\ket{\kv,\sigma} \propto \sum_{a_1,a_2} \ee^{-2\pi \ii (\kappa _1 a_1 + \kappa _2 a_2)}\ket{a_1 \ev _1 + a_2 \ev _2 + \sigma \zetav}\punc{,}
\eeq
where $\kv = \kappa _1 \ev^*_1 + \kappa _2 \ev^*_2$. The nonuniform phase factors $\chi_\Qop(\xv)$ cause the operator $\Qhat$ to mix a discrete set of momentum values, but at certain high-symmetry points in the Brillouin zone (BZ), a smaller set of momenta are mixed.

On the dice lattice with $f = \frac{1}{6}$, one finds that a generic momentum state belongs to a set of $24$ states that are mixed, but that there are four points within the (lattice) BZ that are closed under the action of the full symmetry group. With our choice of gauge field, their momenta are given by $\kv _{m,n} = -(\frac{1}{12}+\frac{m}{2})\ev^*_1 + (\frac{1}{12} + \frac{n}{2})\ev^*_2$ with $m,n \in \{0,1\}$, and we expect the global minima of the single-particle dispersion to occur at these points.\footnote{It is  possible to construct a Hamiltonian $\Ham_1$ so that these are only local extrema, and the global minima of the dispersion are elsewhere. The low-energy sector then involves more than two vortex fields, and the number of possible ordered states is necessarily greater than the six allowed by the microscopic model.}

In fact, we argue that two independent linear combinations from the four points form global minima, as follows: It can be seen from Figure~\ref{FigKagomeDice} that $\KagomeTrans _2$ remains a full symmetry in the presence of the background gauge field, and so the corresponding operator $\hat{K}_2$ takes a particularly simple form. Its action on a position state is given in \refeq{EqChiK2}; acting on a momentum state it gives simply
\beq{EqKagomeTrans2}
\hat{K}_2 \ket{\kv, \sigma} = \ee^{\ii\ev_2 \cdot \kv} \ket{\kv, \sigma}\punc{.}
\eeq
This implies that the single-particle Hamiltonian $\Ham _1$ does not mix momentum states $\ket{\kv_{m,n},\sigma}$ with distinct values of $n$. Considering first $n=0$, one of the energy eigenstates at the minimum of the single-particle dispersion can therefore be written
\beq{EqKet0}
\ket{0} = \sum_{m,\sigma} c_{m\sigma} \ket{\kv_{m,0},\sigma}\punc{,}
\eeq
where $c_{m\sigma}$ are coefficients depending on the details of $\Ham_1$. (The magnetic BZ is half the size of the lattice BZ, so that momenta $\kv _{m,n}$ with $m = 0,1$ correspond to the same point in the magnetic BZ, and are hence mixed by $\Ham _1$.)

The state $\ket{0}$ is clearly an eigenstate of $\hat{K}_2$, with eigenvalue $\ee^{\ii\pi/6}$. It is straightforward to show, using \refeq{EqCommutation1}, that the state $\ket{1} = \ee^{\ii \pi/6} \hat{K}_1 \ket{0}$ is also an eigenstate of $\hat{K}_2$, with eigenvalue $-\ee^{\ii\pi/6}$, and hence $\langle 0 | 1 \rangle = 0$. Since $\hat{K}_1$ commutes with $\Ham_1$ (by construction), this state is also an energy eigenstate with equal eigenvalue. Using the Fourier transform of \refeq{EqChis1}, one finds explicitly
\beq{EqKet1}
\ket{1} = \sum_{m,\sigma} c_{m\sigma} (-1)^m \ket{\kv_{m,1},\sigma}\punc{,}
\eeq
and $\hat{K}_1\ket{1} = \ee^{-\ii \pi/6}\ket{0}$.

There are therefore two degenerate minima of the single-particle dispersion, $\ket{n}$ with $n \in \{0,1\}$. (The same result was found by Jiang and Ye\cite{Jiang} by directly diagonalizing a Hamiltonian with nearest-neighbor hopping.)  While the values of the coefficients $c_{m\sigma}$ depend on the precise form of the vortex hopping Hamiltonian, the behavior of $\ket{0}$ and $\ket{1}$ under the action of the full symmetry group can be determined uniquely using \refeqs{EqCommutation1}{EqCommutation2}, giving
\begin{align}
\label{Eq01Transform1}
\bra{n}\hat{K}_1\ket{n'} &= \ee^{-\ii\pi/6}
\begin{pmatrix}
0 & 1\\
1 & 0
\end{pmatrix}_{nn'}\\
\bra{n}\hat{K}_2\ket{n'} &= \ee^{\ii\pi/6}
\begin{pmatrix}
1 & 0\\
0 & -1
\end{pmatrix}_{nn'}\\
\bra{n}\hat{R}\ket{n'} &=
\frac{\ee^{-\ii\pi/12}}{\sqrt{2}}
\begin{pmatrix}
1&1\\
\ii&-\ii
\end{pmatrix}_{nn'}\\
\bra{n}\hat{X}_1\ket{n'} &=
\frac{1}{\sqrt{2}}
\begin{pmatrix}
1&\ii\\
\ii&1
\end{pmatrix}_{nn'}\punc{.}
\label{Eq01Transform2}
\end{align}

We have so far considered only the single-particle kinetic terms in the action $\Act \sub{dual}$ given in \refeq{EqDualAction}. To return to the full description, we first define creation operators $v_0^\dagger$ and $v_1^\dagger$ for the single particle states $\ket{0}$ and $\ket{1}$. The real-space vortex creation operator projected into the low-energy sector can then be written as
\beq{EqVortexOperator}
\psi^\dagger(\xv) = \varphi _0^* (\xv) v_0^\dagger + \varphi _1^* (\xv) v_1^\dagger\punc{,}
\eeq
where $\varphi _n (\xv)$ are slowly varying functions (on the lattice scale).

The symmetry properties of the two-component vector $\phiv = \begin{pmatrix}\varphi _0\\ \varphi _1\end{pmatrix}$ are determined by those of the states $\ket{n}$, and can be summarized as:
\begin{align}
\label{EqPhiTrans1}
\phiv
&\xrightarrow{\KagomeTrans _1}
\ee^{\ii\pi/6}
\begin{pmatrix}
0 & 1\\
1 & 0
\end{pmatrix}
\phiv\\
\phiv
&\xrightarrow{\KagomeTrans _2}
\ee^{-\ii\pi/6}
\begin{pmatrix}
1 & 0\\
0 & -1
\end{pmatrix}
\phiv\\
\phiv
&\xrightarrow{\Rot}
\frac{\ee^{\ii \pi / 12}}{\sqrt{2}}
\begin{pmatrix}
1&1\\
-\ii&\ii
\end{pmatrix}
\phiv\\
\phiv
&\xrightarrow{\RfX _1}
\frac{1}{\sqrt{2}}
\begin{pmatrix}
1&-\ii\\
-\ii&1
\end{pmatrix}
\phiv^*\punc{.}
\label{EqPhiTrans2}
\end{align}
Note that the reflection $\RfX _1$ is represented by an antiunitary operation.

\subsection{Dual continuum action}

The action $\Act \sub{dual}$ can be rewritten in terms of $\phiv$, and the other components of the vortex fields integrated out, giving renormalized values for the coefficients. The transformation properties of $\phiv$ reproduce the nontrivial effects of the lattice structure and fractional filling $f$, allowing the continuum limit to be taken, with the spatial coordinates $\xv$ extended from discrete values describing the dice lattice to continuous two-dimensional coordinates.

The transformations given in \refeqs{EqPhiTrans1}{EqPhiTrans2} strongly constrain the form of the continuum action, which can be expressed as a series in powers of the vortex field $\phiv$, the gauge field $\Av$, and the space- and time-derivative operators. As in the Coulomb phase, described in Section~\ref{SecCoulombPhase}, the full cubic symmetry of the original dimer problem also constrains the final action to be space-time symmetric.

We first consider the terms containing only the vortex fields $\phiv$. These are most easily found by expressing all gauge-invariant (i.e., local and phase-rotation invariant) bilinears of the fields $\varphi _n$ in terms of the boson density operators, which are the order parameters for the density-wave phases. The ordering patterns with which we are concerned (such as the one illustrated in Figures~\ref{FigCubic111Projection} and \ref{FigCubicKagome3D}) are characterized by nonzero expectation values of the momentum-space density at $\kv \in \{\frac{1}{2}\ev _1^*,\frac{1}{2}\ev _2^*,\frac{1}{2}\ev _1^*+\frac{1}{2}\ev _2^*\}$.

Defining $\rho _{\kappa _1\kappa _2}$ as the boson density operator at momentum $\kv = \kappa _1 \ev _1^* + \kappa _2 \ev _2^*$, the order parameters are $\rho _{\frac{1}{2}0}$, $\rho _{0\frac{1}{2}}$, and $\rho _{\frac{1}{2}\frac{1}{2}}$. In the continuum limit, we can identify these fields with the bilinears of $\varphi _n$ using their symmetry properties; we find
\beq{EqBosonDensity}
\begin{Bmatrix}
\rho_{\frac{1}{2}\frac{1}{2}}\\
\rho_{\frac{1}{2}0}\\
\rho_{0\frac{1}{2}}
\end{Bmatrix}
\sim \phiv^\dagger \Mm^\dagger
\begin{Bmatrix}
\sigv^x\\
\sigv^y\\
\sigv^z
\end{Bmatrix}
\Mm \phiv
\punc{,}
\eeq
where $\Mm = (\boldsymbol{1} - \ii \sigv^x)/\sqrt{2}$ is a unitary matrix, and $\sigv^\mu$ is a Pauli matrix. In the language of the original dimer problem, the order parameter is the magnetization $m _\mu$, related to the dimer occupation number by \refeq{EqOrderParameter}. This can similarly be expressed in terms of the vortex fields as
\beq{EqMagnetizationFromVortices}
m_\mu \sim \phiv^\dagger \Mm^\dagger \sigv^\mu \Mm \phiv\nd\punc{.}
\eeq

We will now identify the allowed interaction terms for the vortex fields, and it is convenient to do this by writing them in terms of $m _\mu$. While the full action has cubic symmetry, the lowest-order terms in fact have \SO3 rotation symmetry, which corresponds, via \refeq{EqMagnetizationFromVortices}, to \SU2 symmetry for the vortex fields $\phiv$. Our primary concern will be to identify the first term that explicitly breaks this higher symmetry.

Defining the \SU2 Casimir invariant $\Omega = \phiv^\dagger \phiv$, all gauge-invariant combinations of the fields $\varphi _n$ can be written in terms of $m _{x,y,z}$ and $\Omega$, and it is straightforward to show that $|\mv|^2 \sim \Omega^2$ is also an \SU2 invariant. While the action can contain any term involving only $\Omega$, terms involving functions of $\phiv$ that break \SU2 are strongly constrained by symmetry.

First note that the only allowed quadratic and quartic (in the vortex fields) combinations are $\Omega$ and $\Omega^2$. At sixth order, besides $\Omega^3$, one finds the combination $m_x m_y m_z$, which is invariant under all of the operations in \refeqs{EqPhiTrans1}{EqPhiTrans2}. It is nonetheless excluded by requiring symmetry under the cubic reflections $\CubicTrans _\mu$, which take $m _\mu \rightarrow -m _\mu$. (In terms of the bosons, this corresponds to particle--hole symmetry, which swaps vortices and antivortices.)

The lowest-order combination that satisfies all the symmetries of the problem, but explicitly breaks the \SU2 symmetry, is of eighth order in $\phiv$, and its contribution to the action is
\beq{EqLag1}
\Lag _1 = v \sum _\mu m _\mu^4\punc{.}
\eeq
Symmetry does not fix the coefficient $v$, and it is in general allowed to take either sign. In the ordered phases that we describe here, however, a single component of the magnetization acquires a nonzero expectation value; such phases require $v < 0$.

\subsection{Emergent \SU2 symmetry}

The full action is given by the continuum limit of \refeq{EqDualAction} and can be written as a three-dimensional integral over a Lagrangian density $\Lag = \Lag _0 + \Lag _1 + \cdots$. The lowest order terms $\Lag _0$ take the forms expected for a \U1 gauge field minimally coupled to matter fields $\phiv$, with the lattice curl becoming a differential curl, and the modified hopping term becoming a covariant derivative:
\beq{EqLag0}
\Lag _0 = |(\del - \ii\Av)\phiv|^2 + s|\phiv|^2 + u(|\phiv|^2)^2 + \kappa |\del \times \Av |^2\punc{.}
\eeq
The interaction term $\Lag _1$ is given in \refeq{EqLag1} and the ellipsis represents further terms that are expected to be irrelevant. The cubic symmetry of the original dimer problem means that $\Lag$ (considered as the action for a quantum problem) is space--time symmetric, and this fact has been used to express $\Lag _0$ in terms of the three-dimensional derivative operator $\del$.

The transition occurs when the quadratic coefficient $s$ is tuned through its critical value, and the gauge field $\Av$ acquires a gap by the Anderson-Higgs mechanism. The superfluid phase of the boson problem (the Coulomb phase of the dimer problem) is equivalent to the Coulomb phase of this gauge theory, within which the long-range correlations are reproduced by the gapless gauge field. The Mott insulator (ordered dimers) corresponds to the Higgs phase, where there are no gapless excitations and \SU2 symmetry is broken by condensation of the matter field $\phiv$.

Note that the gauge field $\Av$ is by construction noncompact, so that monopoles are forbidden. Such monopoles would give points in space-time with nonzero divergence $\divv \Jv$ of the boson current, which map to monomers in the original dimer problem.

It is not firmly established whether the action $\Lag$ has any nontrivial fixed points under the renormalization group (RG). As noted by Balents et al.,\cite{Balents} such a conjecture is difficult to test analytically, since, for example, an expansion in $\epsilon = 4 - d$, where $d$ is the spatial dimension, has no weak-coupling fixed points. As we remark in Section~\ref{SecDiscussion}, the numerical evidence remains inconclusive. It is clear from simulations, however, that the transition in the original dimer model is, at most, weakly first order, with a correlation length of at least hundreds of lattice spacings, and so a continuum description is appropriate.

The terms appearing in $\Lag _0$ have full \SU2 symmetry, while the lowest-order term breaking this to the microscopic cubic symmetry, $\Lag _1$, is of eighth order in the field $\phiv$. It is therefore highly likely that this term, as well as higher-order symmetry-breaking terms, is irrelevant in the continuum, and that the effective theory is given by $\Lag _0$ and has an emergent \SU2 symmetry.

This leads naturally to the conclusion that physical properties measured sufficiently close to the transition should show full \SO3 symmetry, rather than the reduced cubic symmetry of microscopic model. This claim is in fact in agreement with qualitative observations made by Misguich et al.,\cite{Misguich}\ based on their numerical results near the transition. First, the measured distribution of the order parameter $\mv$ at the critically becomes increasingly spherically symmetric for larger system sizes. Second, the dimer correlation function $\langle d_\mu (\rv) d_\nu (\zerov) \rangle$, while taking the form given in \refeq{EqDipolarCorrelations} deep within the Coulomb phase, is dominated by a ``spinlike'' contribution near the transition. This follows from the fact that the magnetization couples directly (without derivatives) to a bilinear in the critical field, according to \refeq{EqMagnetizationFromVortices}, and forms a three-dimensional representation of \SU2. We therefore expect
\beq{EqMagnetizationCorrelations}
\langle m_\mu (\rv) m_\nu (\zerov) \rangle \sim \delta _{\mu\nu}|\rv|^{-d+2-\eta _m}\punc{,}
\eeq
where the critical exponent $\eta _m$ is the anomalous dimension of the magnetization (and a similar expression holds for the dimer correlation function). The absence of a weak-coupling fixed point prevents us from making quantitative predictions about the anomalous dimension $\eta _m$.

As noted by Misguich et al.,\cite{Misguich} these properties, while explained straightforwardly by an \SU2-symmetric continuum theory, are incompatible with other, more obvious, candidate continuum theories, such as the \O3 model.

\section{Discussion}
\label{SecDiscussion}

In this paper we have presented a derivation of a continuum theory to describe the phase transition from a dimer crystal to a Coulomb phase observed in simulations of a classical dimer model on the cubic lattice.\cite{Alet} Our approach proceeds by first mapping to a quantum model of bosons, which has a corresponding phase transition from a Mott insulator with density-wave order to a superfluid. The second stage of the derivation consists of applying a duality mapping to this quantum model, allowing the phase transition to be described in terms of an Anderson-Higgs transition for the dual vortex fields.

The continuum theory we derive coincides with one obtained for the same model by Charrier et al.\cite{Charrier}\ and by Chen et al.,\cite{Chen} using a duality mapping applied directly to the classical model. The Lagrangian $\Lag _0$ is given in \refeq{EqLag0} and is referred to as \NCP; we expect the long-wavelength properties near the transition to be described by this theory. In other words, there should be some range of length scales, much larger than the lattice scale, at which $\Lag _0$ provides an appropriate description. This assumes that $\Lag _1$ and all higher-order terms are irrelevant in the RG sense. As we have noted above, the numerical results of Misguich et al.\cite{Misguich}\ support the claim that at least $\Lag _1$ and all other terms breaking \SU2 symmetry are irrelevant.

The question of whether the transition in the dimer model is continuous or first order remains open, but is clearly related to same question for the transition between the Coulomb and Anderson-Higgs phases in the model $\Lag _0$. This remains contentious,\cite{Motrunich2,Motrunich,Sandvik,Melko,Kuklov,Kuklov2} with numerical evidence so far inconclusive and little prospect of insight from analytics. The theory $\Lag _0$ has one free parameter $\kappa$ (after $s$ is tuned to its critical value), and it is possible that it has a continuous transition for some value of $\kappa$ and not for others.\cite{Motrunich2} (It is not clear how $\kappa$ depends on the details of the classical energy $\clH$---for instance, whether it increases or decreases when one adds further-neighbor interactions.)

In fact, the most economical interpretation of the available numerical results is that the original cubic dimer model (with only nearest-neighbor interactions) is near a tricritical point. Evidence for this comes from the critical exponents in the original dimer model and the results of adding deformations.

The critical exponents reported by Alet et al.\cite{Alet}\ for the cubic dimer model are, as noted by the authors, consistent with those expected generically at a tricritical point (and in fact those seen at the tricritical point in an \NCP{} model by direct simulations\cite{Motrunich2}). They are inconsistent with those for the generic \NCP{} transition as observed in simulations of various models expected to be described by this theory.\cite{Motrunich2,Motrunich,Sandvik,Melko}

The scenario that the cubic dimer model with only nearest-neighbor interactions is near a tricritical point suggests that modifications to the Hamiltonian should drive the system away from this tricritical point, either to a strongly first-order transition or to the generic continuous transition. Recent numerical results show that some perturbations, including ones that preserve the full symmetry\cite{Papanikolaou} and ones that break it,\cite{Chen} can make the transition clearly first order. Confirmation of the tricritical-point scenario would require the demonstration of a line of continuous transitions for some range of parameters.

Charrier et al.\cite{Charrier}\ have studied a model where \refeq{EqClosePacking} is still enforced, but the variables $d_\mu (\rv)$ are allowed to take all integer values (although their simulations treated a dual model). As for the unperturbed dimer model, they found results consistent with a continuous transition, but in this case with exponents in agreement with those reported for \NCP.\cite{Motrunich2,Motrunich,Sandvik,Melko} By introducing an energy cost for larger values of $d_\mu$, it is possible to interpolate continuously between this generalized version and the original dimer model; simulations for a range of values would be a useful test of the tricritical-point scenario.

\begin{acknowledgments}
We thank S. Simon for helpful discussions. The work was supported in part by EPSRC Grant No.\ EP/D050952/1.
\end{acknowledgments}

\appendix*

\section{Models with an intermediate phase}

In their numerical studies, Chen et al.\cite{Chen}\ considered various modifications of the classical configuration energy $\clH$ that break the cubic symmetry of the original model and reduce the number of degenerate ordered states. While certain modifications lead to strongly first-order transitions, others instead cause the appearance of an intermediate phase between the high-temperature Coulomb and low-temperature ordered phases. We will consider one particular case, referred to by Chen et al.\cite{Chen}\ as the `xy' model, and show how the phase diagram can be interpreted in terms of the mapping to an effective quantum Hamiltonian.

\begin{figure}
\putinscaledfigure{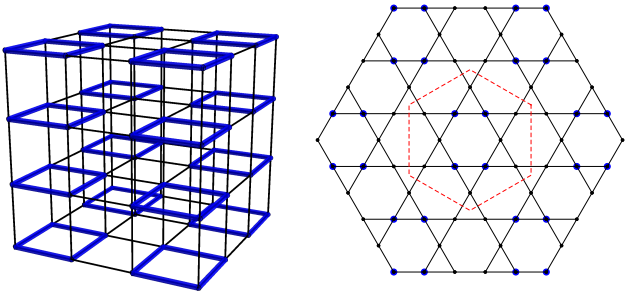}
\caption{\label{FigReducedSymmetry}(color online) Left: Illustration of the `xy' model, in which only dimers on the thick blue bonds contribute to the energy. The cubic symmetry of the original dimer model is reduced and there are only two configurations that minimize the energy. Right: \Kagome\ lattice showing the modifications to the corresponding quantum model; the sites highlighted in blue have an attractive potential energy. The unit cell, shown with dashed red lines, is $4$ times larger than in the case with full symmetry.}
\end{figure}
The `xy' model is defined so that the only configurations that minimize the classical energy $\clH _\text{xy}$ are those with magnetization $\mv \in \{\deltav_x,\deltav_y\}$. This can be achieved by including interactions only between the subset of bonds shown on the left-hand side of Figure~\ref{FigReducedSymmetry}. Explicitly, we can write $\clH _\text{xy} = -(n_x^\mathrm{e} + n_y^\mathrm{e})$, where
\beq{Definenmue}
n_\mu^\mathrm{e} = \sum _{\underset{\text{($r_\mu$ even)}}{\rv}}\sum_{\nu \neq \mu} d_\mu(\rv) d_\mu(\rv + \deltav_\nu)
\eeq
counts the number of nearest-neighbor parallel dimers on even bonds along the $\mu$ direction.

Using the mapping described in Section~\ref{SecMapping}, the bonds included in $\clH _\text{xy}$ correspond to a subset of the sites of the \kagome\ lattice shown on the right-hand side of Figure~\ref{FigReducedSymmetry}. In the quantum Hamiltonian $\quH_\text{xy}$ corresponding to the `xy' dimer model, these sites acquire an attractive potential energy, reducing the degeneracy of the density-wave ordered Mott insulator states that occur in the limit of weak hopping.

Crucially, the reduced symmetry increases the size of the unit cell fourfold to $12$ sites, so that there are now $2$ bosons per unit cell. It is therefore possible to form a zero-temperature insulating state that breaks no symmetries. A simple caricature of such a state has one boson localized on every neighboring pair of attractive sites, forming a valence bond linking the two. The true ground state will involve both fluctuations onto other sites and correlations between the positions of nearby bosons.

Such a valence-bond Mott insulator is expected to be the ground state at intermediate ratios of the hopping strength to the interactions. When the hopping is increased, the fluctuations grow larger, leading to a transition into a superfluid phase with broken \U1 symmetry and gapless Goldstone modes. This transition is described by the standard fixed-density (relativistic) \U1 critical theory, as for the Bose-Hubbard model at integer filling,\cite{Fisher,Sachdev} in the XY universality class. For smaller hopping strengths, the further-neighbor interactions between bosons will become sufficient to cause long-range order in the positions of the bosons. This breaks the symmetry $\RfX_1$ (see Figure~\ref{FigKagomeSymmetries}) giving a density-wave ordered state corresponding to one of the dimer crystals with magnetization $\mv \in \{\deltav_x,\deltav_y\}$. This transition is, if continuous, described by the Ising universality class.

The observations of Chen et al.\cite{Chen}\ are consistent with our analysis. At high temperatures, they observe a Coulomb phase (corresponding to the quantum superfluid), as in the original model with full cubic symmetry. As the temperature is lowered, the system enters a `paramagnetic' phase (the valence-bond Mott insulator), with no broken symmetry and confined monomers, via a transition in the inverted-XY class. Symmetry between the $x$ and $y$ directions ($\RfX_1$ on the \kagome\ lattice) is broken at a lower-temperature transition into the dimer crystal (quantum density-wave order), which is in fact found to be strongly first-order. By considering the interactions that can be added to the Lagrangian $\Lag_0$ [\refeq{EqLag0}] in this reduced-symmetry case, Chen et al.\cite{Chen}\ have shown that this first-order transition can be naturally understood as a ``spin flop'', a reorientation of the vector order parameter $\phiv$.


\begin{thebibliography}{99}

\bibitem{Youngblood} R. Youngblood, J. D. Axe, and B. M. McCoy, Phys.\ Rev.\ B {\bf 21}, 5212 (1980).

\bibitem{Henley} C. L. Henley, Can.\ J. Phys.\ {\bf 79}, 1307 (2001).

\bibitem{Huse} D. A. Huse, W. Krauth, R. Moessner, and S. L. Sondhi, Phys.\ Rev.\ Lett.\ {\bf 91}, 167004 (2003).

\bibitem{Bramwell} S. T. Bramwell and M. J. P. Gingras, Science {\bf 294}, 1495 (2001).

\bibitem{Kasteleyn} P. W. Kasteleyn, J. Math.\ Phys.\ {\bf 4}, 287 (1963).

\bibitem{Blunt} M. O. Blunt, J. C. Russell, M. C. Gim\'enez-L\'opez, J. P. Garrahan, X. Lin, M. Schr\"oder, N. R. Champness, and P. H. Beton, Science {\bf 322}, 1077 (2008).

\bibitem{Landau} L. D. Landau and E. M. Lifshitz, {\em Statistical Physics} (Butterworth-Heinemann, New York, 1999).

\bibitem{Bergman} D. L. Bergman, G. A. Fiete, and L. Balents, Phys.\ Rev.\ B {\bf 73}, 134402 (2006).

\bibitem{Alet} F. Alet, G. Misguich, V. Pasquier, R. Moessner, and J. L. Jacobsen, Phys.\ Rev.\ Lett.\ {\bf 97}, 030403 (2006).

\bibitem{Jaubert} L. D. C. Jaubert, J. T. Chalker, P. C. W. Holdsworth, and R. Moessner, Phys.\ Rev.\ Lett.\ {\bf 100}, 067207 (2008).

\bibitem{SpinIce} S. Powell and J. T. Chalker, Phys.\ Rev. B {\bf 78}, 024422 (2008).

\bibitem{Letter} S. Powell and J. T. Chalker, Phys.\ Rev.\ Lett.\ {\bf 101}, 155702 (2008).

\bibitem{Pickles} T. S. Pickles, T. E. Saunders, and J. T. Chalker, Europhys.\ Lett.\ {\bf 84}, 36002 (2008).

\bibitem{Fowler} R. H. Fowler and G. S. Rushbrooke, Trans.\ Faraday Soc.\ {\bf 33}, 1272 (1937).

\bibitem{Misguich} G. Misguich, V. Pasquier, and F. Alet, Phys.\ Rev.\ B {\bf 78}, 100402(R) (2008).

\bibitem{Charrier} D. Charrier, F. Alet, and P. Pujol, Phys.\ Rev.\ Lett.\ {\bf 101}, 167205 (2008).

\bibitem{Chen} G. Chen, J. Gukelberger, S. Trebst, F. Alet, and L. Balents, Phys.\ Rev.\ B {\bf 80}, 045112 (2009).

\bibitem{Papanikolaou} S. Papanikolaou and J. J. Betouras, arXiv:0904.0272 (unpublished).

\bibitem{Balents} L. Balents, L. Bartosch, A. Burkov, S. Sachdev, and K. Sengupta, Phys.\ Rev.\ B {\bf 71}, 144508 (2005); Prog.\ Theor.\ Phys.\ Suppl.\ {\bf 160}, 314 (2005).

\bibitem{Hatano} N. Hatano and D. R. Nelson, Phys.\ Rev.\ B {\bf 56}, 8651 (1997).

\bibitem{Sengupta} K. Sengupta, S. V. Isakov, and Y. B. Kim, Phys.\ Rev.\ B {\bf 73}, 245103 (2006).

\bibitem{Wallin} M. Wallin, E. S. S\o rensen, S. M. Girvin, and A. P. Young, Phys.\ Rev.\ B {\bf 49}, 12115 (1994).

\bibitem{Vidal} J. Vidal, P. Butaud, B. Dou\c cot and R. Mosseri, Phys. Rev. B 64, 155306 (2001).

\bibitem{Jiang} L. Jiang and J. Ye, J. Phys.: Condens.\ Matter {\bf 18}, 6907 (2006).

\bibitem{Wen} X.-G. Wen, Phys.\ Rev.\ B {\bf 65}, 165113 (2002).

\bibitem{Motrunich} O. I. Motrunich and A. Vishwanath, Phys.\ Rev.\ B {\bf 70}, 075104 (2004).

\bibitem{Sandvik} A. W. Sandvik, Phys.\ Rev.\ Lett.\ {\bf 98}, 227202 (2007).

\bibitem{Melko} R. G. Melko and R. K. Kaul, Phys.\ Rev.\ Lett.\ {\bf 100}, 017203 (2008).

\bibitem{Kuklov} A. B. Kuklov, M. Matsumoto, N. V. Prokof'ev, B. V. Svistunov, and M. Troyer, Phys.\ Rev.\ Lett.\ {\bf 101}, 050405 (2008).

\bibitem{Motrunich2} O. I. Motrunich and A. Vishwanath, arXiv:0805.1494v1.

\bibitem{Kuklov2} A. Kuklov, M. Matsumoto, N. Prokof'ev, B. Svistunov, M. Troyer, arXiv:0805.2578v1.

\bibitem{Fisher} M. P. A. Fisher, P. B. Weichman, G. Grinstein, and D. S. Fisher, Phys.\ Rev.\ B {\bf 40}, 546 (1989).

\bibitem{Sachdev} S. Sachdev, {\em Quantum Phase Transitions}, Cambridge University Press, Cambridge (1999).

\end{thebibliography}
\end{document}